\documentclass[twocolumn, secnumarabic, nofootinbib, superscriptaddress, aps, prl]{revtex4-1}

\usepackage{graphicx}%
\usepackage{multirow}
\usepackage{float}
\usepackage{epstopdf}
\usepackage[utf8]{inputenc}
\usepackage[english]{babel}
\usepackage{footnote}
\usepackage{comment}
\usepackage{tensor}
\usepackage{xcolor}
\usepackage{verbatim}

\begin{document}
\title{First direct $^{7}$Be electron capture $Q$-value measurement towards high-precision BSM neutrino physics searches}

\author{R. Bhandari}
\affiliation{Department of Physics, Central Michigan University, Mount Pleasant, Michigan, 48859, USA}

\author{G. Bollen}
\affiliation{Facility for Rare Isotope Beams, East Lansing, Michigan, 48824, USA}
\affiliation{Department of Physics and Astronomy, Michigan State University, East Lansing, Michigan 48824, USA}

\author{T. Brunner}
\affiliation{Department of Physics, McGill University, Montr\'{e}al, Qu\'{e}bec H3A 2T8, Canada}

\author{N. D. Gamage}
\affiliation{Facility for Rare Isotope Beams, East Lansing, Michigan, 48824, USA}

\author{A. Hamaker}
\affiliation{Facility for Rare Isotope Beams, East Lansing, Michigan, 48824, USA}
\affiliation{Department of Physics and Astronomy, Michigan State University, East Lansing, Michigan 48824, USA}

\author{Z. Hockenbery}
\affiliation{Department of Physics, McGill University, Montr\'{e}al, Qu\'{e}bec H3A 2T8, Canada}
\affiliation{TRIUMF, Vancouver, British Columbia V6T 2A3, Canada}

\author{M. Horana Gamage}
\affiliation{Department of Physics, Central Michigan University, Mount Pleasant, Michigan, 48859, USA}

\author{D. K. Keblbeck}
\affiliation{Department of Physics, Central Michigan University, Mount Pleasant, Michigan, 48859, USA}

\author{K. G. Leach}
\affiliation{Department of Physics, Colorado School of Mines, Golden, Colorado 80401, USA}
\affiliation{Facility for Rare Isotope Beams, East Lansing, Michigan, 48824, USA}

\author{D. Puentes}
\affiliation{Facility for Rare Isotope Beams, East Lansing, Michigan, 48824, USA}
\affiliation{Department of Physics and Astronomy, Michigan State University, East Lansing, Michigan 48824, USA}

\author{M. Redshaw}
\affiliation{Department of Physics, Central Michigan University, Mount Pleasant, Michigan, 48859, USA}
\affiliation{Facility for Rare Isotope Beams, East Lansing, Michigan, 48824, USA}

\author{R. Ringle}
\affiliation{Facility for Rare Isotope Beams, East Lansing, Michigan, 48824, USA}

\author{S. Schwarz}
\affiliation{Facility for Rare Isotope Beams, East Lansing, Michigan, 48824, USA}

\author{C. S. Sumithrarachchi}
\affiliation{Facility for Rare Isotope Beams, East Lansing, Michigan, 48824, USA}

\author{I. Yandow}
\affiliation{Facility for Rare Isotope Beams, East Lansing, Michigan, 48824, USA}
\affiliation{Department of Physics and Astronomy, Michigan State University, East Lansing, Michigan 48824, USA}

\date{\today}%

\begin{abstract}
\noindent

We report the first direct measurement of the nuclear electron capture (EC) decay $Q$-value of $^{7}$Be $\rightarrow$ $^{7}$Li via high-precision Penning trap mass spectrometry (PTMS).  This was performed using the LEBIT Penning trap located at the National Superconducting Cyclotron Laboratory/Facility for Rare Isotope Beams (NSCL/FRIB) using the newly commissioned Batch-Mode Ion-Source (BMIS) to deliver the unstable $^{7}$Be$^{+}$ samples.  With a measured value of $Q_{EC}$ = 861.963(23) keV this result is three times more precise than any previous determination of this quantity.  This improved precision, and accuracy of the $^7$Be EC decay $Q$-value is critical for ongoing experiments that measure the recoiling nucleus in this system as a signature to search for beyond Standard Model (BSM) neutrino physics using $^7$Be-doped superconducting sensors. This experiment has extended LEBIT capabilities, using the first low-energy beam delivered by BMIS at FRIB for PTMS, as well as measuring the lightest-mass isotopes so far with LEBIT.

\end{abstract}

\maketitle

The experimental observation of neutrino oscillations has provided the only known evidence of deviation from the Standard Model (SM) description of the known fundamental particles---non-zero neutrino mass states~\cite{Fukuda1998_v,Ahmad2001_SNO}. This fact makes extensions to the SM unavoidable, and at the most basic level, requires any new theory to incorporate neutrino mass and explain its origin. Several well-motivated extensions to the SM include the possibility of additional heavy neutrino mass states that are associated with so-called ``sterile" flavor states that are even more weakly coupled to the SM than the known neutrinos~\cite{deGouvea_nu_mass_review,Dasgupta:2021ies_sterile_review}. Observation of these neutrino mass states would provide a clear path towards a ``new" SM description of neutrinos, and may also help address the dark matter and baryon asymmetry problems of our Universe~\cite{dodelson1994sterile,shaposhnikov2006numsm,Boyarsky:2018tvu_sterile_dm_review}.

Since neutrinos are neutral, weakly interacting particles, direct measurements of their properties are challenging due to the extremely small interaction probabilities.
As a result, clever indirect methods that exploit energy and momentum conservation in nuclear electron capture (EC) decay can be used as precise probes of the neutrino~\cite{SHROCK1980159_kinem,PhysRevD.46.R888_kinem,PhysRevC.58.2512_ar37,2021QS&T....6b4008M_hunter_qst, 2019NJPh...21e3022S_hunter_smith,Friedrich2021_BeEST}.  In this approach, the recoil energy of the final-state atom that is given a momentum ``kick'' from the neutrino following EC decay is measured, and any missing momentum from the known decay $Q$-value is a signature of physics beyond the SM (BSM). Since there is only one way that two massive bodies can share the decay energy ($Q$-value), a high-precision measurement of the daughter atom recoil energy, $T_{D}$, is connected to the mass of the emitted neutrino, $m_{\nu}$, via
\begin{equation}
    T_D = \frac{Q^2_{EC} - m_\nu^2c^4}{2(Q_{EC} + m_Dc^2)},
    \label{Eqn_Erecoil}
\end{equation}
where $m_{D}$ is the mass of the daughter atom. The Beryllium Electron Capture in Superconducting Tunnel Junctions (BeEST) experiment employs this concept using $^{7}$Be implanted in superconducting tunnel junction (STJ) sensors to measure the $^{7}$Li kinetic energy~\cite{Leach2022_BeEST}. The light $^{7}$Be -- $^{7}$Li system, with large $Q_{EC}$ = 861.89(7) keV~\cite{huang2021ame}, results in a relatively large daughter recoil energy, $T_{D}$($^{7}$Li) = 56.826(9) eV, which is well-suited to studies with STJs that have a full width at half-maximum energy resolution of a few eV in the energy range 20 -- 120 eV~\cite{Ponce2018_STJs, Fretwell2020_BeEST}. Furthermore, STJs can be calibrated via multiphoton absorption with a pulsed laser source to a statistical precision of 1 meV~\cite{Friedrich2020_LTD}, potentially opening new precision tests of the SM. The interpretation of any BSM signatures in the BeEST experiment requires a precise and accurate determination of $Q_{EC}$, which is best achieved through direct Penning trap mass spectrometry (PTMS) measurements of $^{7}$Be and $^{7}$Li ions.

The $^{7}$Be $Q$-value listed in the most recent atomic mass evaluation (AME2020)~\cite{huang2021ame} with a precision of 70 eV/$c^{2}$ is obtained from the energy equivalent of the mass difference between parent and daughter atoms,
\begin{equation}
    Q_{EC} = [M(^{7}\mathrm{Be}) - M(^{7}\mathrm{Li})]c^{2}.
    \label{Eqn_QEC}
\end{equation}
The mass of $^{7}$Li has been measured using Penning trap mass spectrometry to a precision of 4 eV/$c^{2}$~\cite{Nagy2006_7Li}. The mass of $^{7}$Be on the other hand is known to only 70 eV/$c^{2}$, and is determined from four $^{7}$Li($p,n$)$^{7}$Be reaction measurements performed in the 1960s -- 1980s~\cite{Rytz1961_7Li,Gasten1963_7Li,Roush1970_7Li,White1985_7Li}, and never previously by PTMS. The $Q_{EC}$ value has also never been measured directly via the mass difference of parent and daughter atoms. In this Letter we report the first direct PTMS measurement of the $^{7}$Be mass, and the first direct $Q_{EC}$ determination from a measurement of the $^{7}$Be$^{+}$/$^{7}$Li$^{+}$ mass ratio.

\textit{Methods}---The $^{7}$Be EC $Q$-value measurement was performed with the Low Energy Beam and Ion Trap (LEBIT) Penning trap mass spectrometry facility during the transition period between laboratory operations as the National Superconducting Cyclotron Laboratory (NSCL) and the Facility for Rare Isotope Beams (FRIB). The $Q$-value was determined from a measurement of the cyclotron frequency ratio of $^{7}$Be$^{+}$ and $^{7}$Li$^{+}$ ions in the Penning trap, as described below. These measurements extend the reach of LEBIT to the lightest isotopes to which it has been applied. They also utilize for the first time the capability of the recently commissioned Batch Mode Ion Source (BMIS)~\cite{Sumithrarachchi2023_BMIS} for a Penning trap measurement.

\begin{figure}[t]
 \centering
    \includegraphics[width=0.5\textwidth]{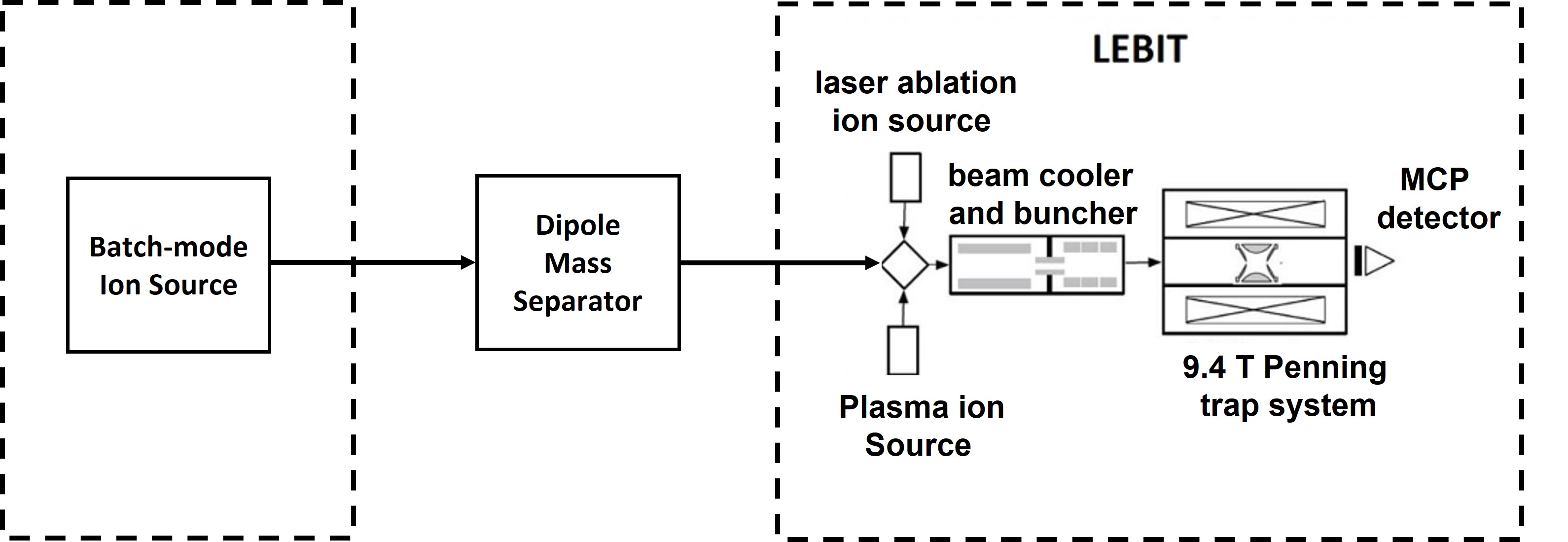}
    \caption{Schematic layout of the batch mode ion source and LEBIT facility connected via the transfer line following a dipole mass separator. The apparatus inside the dashed boxes are held on a 30 kV platform to facilitate ion transport from the ion source to LEBIT.}
   \label{Fig:LEBIT}
\end{figure}

A schematic of the LEBIT facility and other components relevant to this measurement is shown in Fig.~\ref{Fig:LEBIT}. A beam of the 53 day half-life $^{7}$Be isotope was produced with the BMIS, analyzed by a dipole mass separator, and delivered to LEBIT as singly-charged ions. Two separate $^{7}$Be sources were used during the course of this measurement with activities of 1.6 mCi and 4.6 mCi, which are referred to as Run I and Run II below.
Singly-charged ions of the daughter isotope, $^{7}$Li, were produced with the LEBIT laser ablation ion source (LAS)~\cite{izzo2016laser} in which a 25 mm $\times$ 25 mm $\times$ 0.6 mm thick sheet of naturally abundant, 99.9\% purity lithium was installed~\cite{Goodfellow}. 
Once ions from either the BMIS or LAS enter the main LEBIT beamline they encounter the beam cooler buncher~\cite{Schwarz2016}, which produces low emittance pulsed beams that are then ejected and travel toward the LEBIT Penning trap, housed inside a 9.4 T superconducting solenoidal magnet~\cite{Ringle2009_LEBIT}. In this experiment, the time-of-flight ion cyclotron frequency resonance (TOF-ICR) technique~\cite{Gra1980,konig1995} was used to measure the cyclotron frequency of the $^{7}$Be$^{+}$ or $^{7}$Li$^{+}$ ions. Briefly, ions are captured in the Penning trap on a magnetron orbit with radius $\approx$1 mm, created by steering the ions away from the trap center with a ``Lorentz steerer'' just before they enter it~\cite{Ringle2007_Lorentz}. The ions are then subjected to a radiofrequency (rf) quadrupolar electric field applied across the segmented ring electrode for a time $T_{rf}$. The rf is applied at a frequency $\nu_{rf} \approx \nu_{c}$, where
\begin{equation}
    \nu_{c} = \frac{qB}{2\pi m}
    \label{Eqn:fc}
\end{equation}
is the true cyclotron frequency for an ion with mass to charge ratio $m/q$ in a uniform magnetic field of strength $B$.
When $\nu_{rf} = \nu_{+} + \nu_{-}$, magnetron motion, with frequency $\nu_{-}$, is optimally converted into cyclotron motion, with frequency $\nu_{+}$. The value of $\nu_{rf}$ at this resonant condition is taken as $\nu_{c}$ based on the relationship
\begin{equation}
    \nu_{c} = \nu_{+} + \nu_{-},
    \label{Eqn:fc+-}
\end{equation}
which is true for an ideal Penning trap, and can be shown to be valid for a real Penning trap to an accuracy well below the statistical precision achieved here~\cite{Gabrielse2009_sideband,Gabrielse2009_IJMS_Sidebands}.

%
%

\begin{figure}[h]
 \centering
    \includegraphics[width=0.4\textwidth]{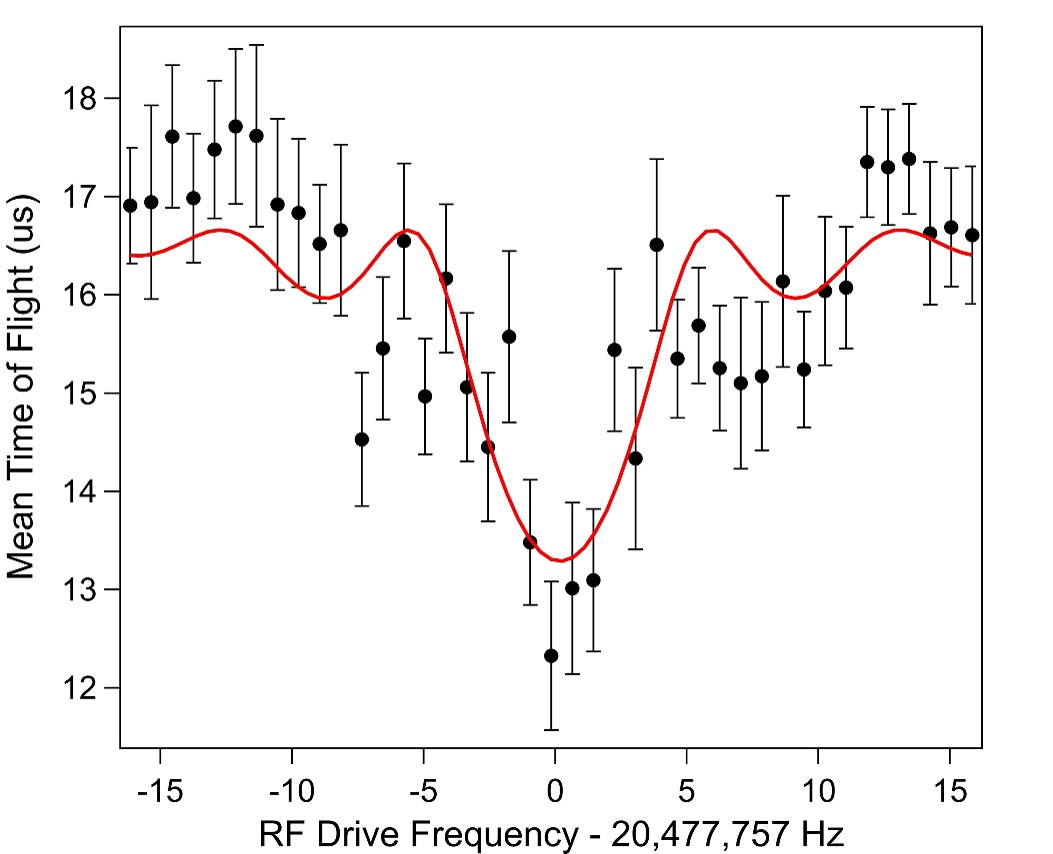}
    \caption{(color online) Time-of-flight cyclotron frequency resonance for $^{7}$Be$^{+}$ using a 150 ms excitation time. The solid red line is a fit to the theoretical line shape~\cite{konig1995}.}
   \label{FIG:TOF_Resonance}
\end{figure}

Next, ions are ejected from the trap and travel toward a microchannel plate (MCP) detector. The TOF is reduced for ions with more radial energy i.e. a larger cyclotron amplitude in the trap. Hence, a minimum in TOF occurs when $\nu_{rf} = \nu_{c}$. The measurement procedure involves capturing a bunch of typically 1 -- 5 ions in the Penning trap, applying the quadrupole rf pulse at a frequency close to $\nu_{c}$, ejecting the ions from the trap, and measuring their TOF. This scheme is repeated while systematically varying $\nu_{rf}$. Hence, a TOF resonance is built up. An example of data from a single $T_{rf}$ = 150 ms excitation time TOF resonance is shown in Fig.~\ref{FIG:TOF_Resonance}. A fit of the theoretical line shape~\cite{konig1995} is applied to the data and the frequency corresponding to the minimum TOF is obtained as a measurement of $\nu_{c}$. In this experiment a typical $^{7}$Be$^{+}$($^{7}$Li$^{+}$) TOF resonance contained $\approx$300 -- 400(1400) ions, took 25(15) minutes, and allowed a $\nu_{c}$ determination to a precision of $\approx$0.4(0.2) Hz. The main limitations on the statistical precision achieved were the measurement time and contaminant ions. The measurement time was limited to 150 ms due to the increased damping effects for the low mass/high frequency ions used here. Contaminant ions were cleaned with rf dipole drive pulses at their respective $\nu_{+}$ frequencies. However, cleaning is never 100 \% efficient and contaminants ions that are detected on the MCP reduce the TOF effect of the resonant ions, making the statistical precision worse. As discussed below, the low rate of contaminant ions did not result in systematic frequency shifts.

In order to determine the cyclotron frequency ratio, $R$, of $^{7}$Be$^{+}$ and $^{7}$Li$^{+}$, corresponding to the inverse mass ratio of ions,
\begin{equation} \label{Eqn1}
 R = \frac{\nu_c(^7\textrm{Be}^+)}{\nu_c(^7\textrm{Li}^+)}=\frac{m(^7\textrm{Li}^+)}{m(^7\textrm{Be}^+)},
 \label{Eqn:CFR}
\end{equation}
we alternately performed $\nu_{c}$ measurements like the one shown in Fig.~\ref{FIG:TOF_Resonance} on $^{7}$Li$^{+}$ and $^{7}$Be$^{+}$. As such, two $\nu_c$($^{7}$Li$^{+}$) measurements enclose each $\nu_c$($^{7}$Be$^{+}$) measurement. Each pair of $\nu_c$($^{7}$Li$^{+}$) measurements were linearly interpolated to find $\nu_c$($^{7}$Li$^{+}$) at the time of the $\nu_c$($^{7}$Be$^{+}$) measurement to account for linear magnetic field drifts. The effect of non-linear field drifts has been previously investigated for the LEBIT system and shown to affect $R$ at the level of $\leq$10$^{-9}$ per hour, which, for the measurement time and statistical uncertainty of an individual $\nu_c$ measurement provides a negligible contribution~\cite{Ringle2007_37_38Ca}.

\begin{figure}[h]
 \centering
    \includegraphics[width=0.5\textwidth]{
    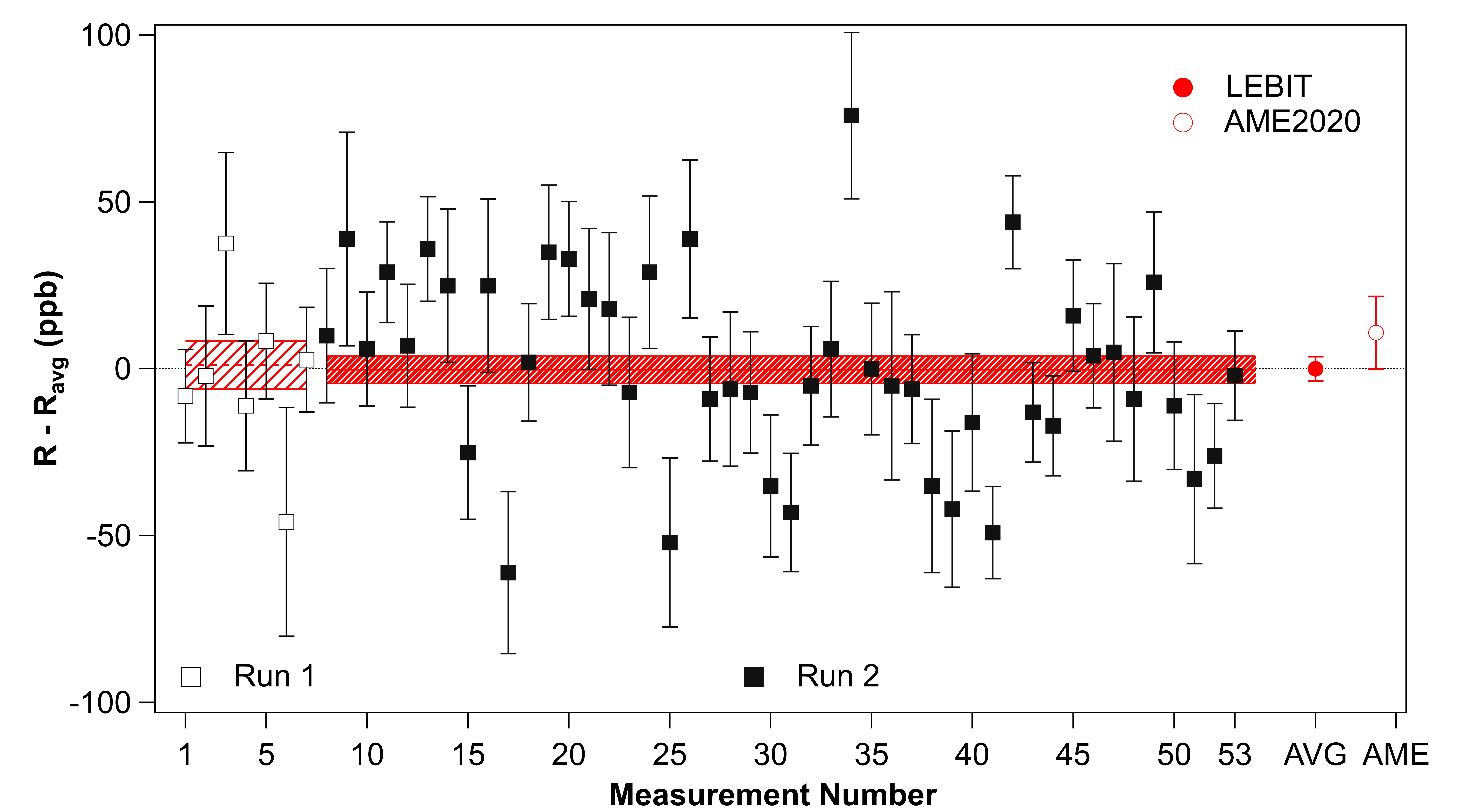}
    \caption{(color online) Difference in individual cyclotron frequency ratio measurements from Run I (open squares) and II (solid squares), compared to the average value, $R_{\mathrm{avg}}$ of Run I and II, respectively, as listed in Table \ref{table:ratio}. The light (heavy) shading indicates the $\pm1\sigma$ uncertainty on $R_{\mathrm{avg}}$ for Run I (II). The solid circle data point represents our final average and uncertainty from Run I and II combined, and the open circle represents the ratio obtained using data from AME2020~\cite{huang2021ame} in comparison to our final result.}
   \label{Fig:RatioData}
\end{figure}

During this measurement campaign we performed two experimental runs using two different $^{7}$Be sources. These consisted of 7 and 46 individual cyclotron frequency ratio measurements for Run I and II, respectively. The individual ratio measurements are shown in Fig.~\ref{Fig:RatioData}. The weighted average, $\bar{R}$, and associated statistical uncertainty were obtained and are also shown in Fig.~\ref{Fig:RatioData}. To evaluate how well the individual statistical uncertainties describe the distribution of measurements of $R$, we determined the Birge ratio~\cite{Birge1932}, which is expected to be $\approx$1. If the Birge ratio was found to be $>$1, the corresponding statistical uncertainty was inflated by the Birge ratio.

\textit{Results and Discussion}---The average cyclotron frequency ratios that we obtained for the two data sets are listed in Table \ref{table:ratio}, along with their weighted average. A statistical precision of 3.6 $\times$ 10$^{-9}$ in the final cyclotron frequency ratio was obtained. We also considered potential sources of systematic uncertainty that included frequency shifts due to (i) the Coulomb interaction between ions in the trap, (ii) the effect of deviations from a perfectly uniform magnetic field or perfectly quadratic electrostatic potential in the trap, and (iii) the effect of relativistic mass increase. The latter two shifts depend on the normal mode amplitudes for ions in the trap, and can be significant for individual ions, but typically cancel in the cyclotron frequency ratio when comparing ions of the same nominal $m/q$, assuming the normal mode amplitudes are the same for both ions. This assumption is expected to hold for ions with the same $m/q$ because they have the same initial conditions in the cooler/buncher and their trajectory to the trap should be the same. 
\begin{table}[t]
\caption{\label{table:ratio} Average cyclotron frequency ratio of $^{7}$Be$^{+}$ vs $^{7}$Li$^{+}$ for the two experimental runs, and their weighted average. $N$ is the number of individual ratio measurements in each run that contributed to the average, $\bar{R}$. The statistical uncertainties are shown in parentheses and have been inflated by the Birge Ratio, BR, when BR $>$ 1.}
\begin{ruledtabular}
\begin{tabular}{ccccc}
Run & Ion Pair & $N$ & BR & $\bar{R}$ \\
\hline
I & $^{7}\text{Be}^+$/$^{7}\text{Li}^+$ & 7 & 0.83 & $0.999\ 868\ 115\ 5(72)$ \\
II & $^{7}\text{Be}^+$/$^{7}\text{Li}^+$ & 46 & 1.45 & $0.999\ 868\ 114\ 1(41)$ \\
Avg. & $^{7}\text{Be}^+$/$^{7}\text{Li}^+$ &  &  & $0.999\ 868\ 114\ 4(36)$ \\
\end{tabular}
\end{ruledtabular}
\end{table}

To investigate (i), we performed a count rate class analysis~\cite{Kellerbauer2003} where we used the fact that our data contained a Poisson distribution for the number of ions, $n_{ion}$, per cycle in the trap. We could therefore determine $\bar{R}$ as a function of $n_{ion}$. From this analysis we found no statistically significant effect on $\bar{R}$ due to $n_{ion}$. Furthermore, we restricted our final analysis to data with $n_{ion} \leq 5$. 

To investigate (ii) and (iii), we took additional data for $^{6}$Li$^{+}$/$^{7}$Li$^{+}$, where $^{6}$Li$^{+}$ ions were also produced from the lithium foil with the LAS. We took data using the same system settings as we did for the $^{7}$Li$^{+}$/$^{7}$Be$^{+}$ measurement, and we used two settings that applied less steering with the Lorentz steerer, placing the ions on a smaller initial magnetron orbit. Hence, we obtained data for $R_{6/7}$ = $\nu_{c}$($^{6}$Li$^{+}$)/$\nu_{c}$($^{7}$Li$^{+}$) vs radial amplitude, $\rho$. Previous studies with the LEBIT apparatus on higher $m/q$ ions where the relativistic shift is negligible, found that the shift due to comparing ions of different $m/q$ was 2 -- 5 $\times$ 10$^{-10}$~\cite{Gulyuz2015_96Zr,Horana2022_75As}, which is small compared to the statistical uncertainty obtained in our current measurements. Therefore, effect (ii) is expected to be small even for $^{6}$Li$^{+}$/$^{7}$Li$^{+}$.

From Eqn. (\ref{Eqn:fc}), the cyclotron frequency shift due to relativistic mass increase is, to lowest order
\begin{equation}
    \frac{\Delta\nu_{c}}{\nu_{c}} \approx \frac{2\pi^{2}\nu_{c}^{2}}{c^2}\rho^{2}.
\end{equation}
Hence, there are two contributions of this shift to the ratio: 1) if ions of different $m/q$ and therefore different $\nu_{c}$ are compared, and 2) if the ions do not have the same value for $\rho$. 

Based on experimental and simulated analyses of the mass dependence of the radial amplitude of ions in the trap when placed on an initial magnetron orbit using the Lorentz steerer~\cite{Ringle2007_Lorentz}, we expected an $\approx$2 \% difference in $\rho$ for $^{6}$Li$^{+}$ and $^{7}$Li$^{+}$, which is small compared to the $\approx$15 \% difference in $\nu_{c}$ due to the difference in $m/q$. Therefore, the shift to $R_{6/7}$ should go as
\begin{equation}
\Delta R_{6/7} \approx \frac{2\pi^{2}}{c^2}\Delta\nu_{c}^{2}\bar{\rho}^{2},
\label{Eqn:DR_6/7}
\end{equation}
where $\Delta\nu_{c}^{2} = \nu_{c}^{2}(^{6}\mathrm{Li}^{+}) - \nu_{c}^{2}(^{7}\mathrm{Li}^{+})$, and $\bar{\rho}$ is the average radial amplitude for $^{6,7}$Li$^{+}$. In our data, we were able to verify a $\Delta R_{6/7} = k\bar{\rho}^{2}$ dependence. For the settings used in our $^{7}$Li$^{+}$/$^{7}$Be$^{+}$ measurement, we observed a $\Delta R_{6/7} \approx3 \times 10^{-8}$ shift, corresponding to an $\approx$200 eV shift in the mass of $^{6}$Li compared to the literature value~\cite{Mount2010_Alkalis} when using $^{7}$Li as a reference. Assuming that  $k = (2\pi^{2}/c^{2})\Delta\nu_{c}^{2}$, our $^{6}$Li$^{+}$/$^{7}$Li$^{+}$ data provided a value for $\bar{\rho} \approx$ 1 mm for the setting used in the $^{7}$Li$^{+}$/$^{7}$Be$^{+}$ data as expected. 
From this analysis, we conclude that, for our $^{7}$Li$^{+}$/$^{7}$Be$^{+}$ ratio measurement, where the fractional mass difference between the two ions is $\sim$1000 times smaller than for $^{6}$Li$^{+}$/$^{7}$Li$^{+}$, the systematic shift due to special relativity and trap imperfections is $\leq 1 \times 10^{-10}$ and is negligible. This corresponds to a shift of $\leq$ 1 eV in the $Q$-value.

Using the value for $\bar{R}$ listed in Table \ref{table:ratio}, we obtain the $Q$-values shown in Table \ref{Table:Q_values} from
\begin{equation}
     Q_{EC} = \left[M(^7\textrm{Li})-m_e\right](\bar{R}^{-1}-1)c^2 - \left(B_{\mathrm{Be}} - B_{\mathrm{Li}}\right).
\end{equation}
$M$($^7\textrm{Li}$) is the atomic mass of $^{7}$Li from AME2020~\cite{huang2021ame}, $m_{e}$ is the mass of the electron~\cite{CODATA2018}, and $B_{\mathrm{Li}}$ = 5.4 eV, $B_{\mathrm{Be}}$ = 9.3 eV are the first ionization energies of lithium and beryllium~\cite{NIST_ASD}, respectively, and must be accounted for at the level of precision achieved here$\footnote{Note, we have used the fact that $\bar{R} \approx 1$}$. Our final result for the $^{7}$Be EC decay $Q$-value is $Q_{EC}$($^{7}$Be) = 861.963(23) keV. The value obtained using AME2020 data agrees with our result at the level of 1$\sigma$, but our new direct measurement is a factor of 3 more precise.
Using our new $Q$-value and Eqn. (\ref{Eqn_Erecoil}), we obtain $T_{D}$ = 56.836(3) eV. 

We were also able to obtain a more precise value for the mass excess of $^{7}$Be from our measurement via
\begin{equation}
     \mathrm{ME(^{7}Be}) = Q_{EC}/c^2 + \mathrm{ME(^{7}Li}).
\end{equation}
Using $\mathrm{ME(^{7}Li})$ = 14\,907.1046(42) keV/$c^2$ from AME2020~\cite{huang2021ame}, we obtain $\mathrm{ME(^{7}Be})$ = 15\,769.067(23) keV/$c^2$. As with the $^{7}$Be $Q_{EC}$-value, our new mass excess is larger than the AME value by 70 eV, but they agree at the 1$\sigma$ level.

\begin{table}[H]
\caption{\label{table:mass} $^{7}$Be $Q_{EC}$-values obtained in this work ($Q_{\rm{LEBIT}}$) and comparison with the value from the AME2020 ($Q_{\rm{AME}}$)~\cite{huang2021ame} where $\Delta Q$ = $Q_{\rm{LEBIT}}$ - $Q_{\rm{AME}}$}
\begin{ruledtabular}
\begin{tabular}{ccccc}
\multirow{2}{*}{Run} &  \multicolumn{1}{c}{This work} & \multicolumn{1}{c}{AME2020} & \multicolumn{1}{c}{$\Delta$Q}\\
   & \multicolumn{1}{c}{$Q_{\rm{LEBIT}}$ (keV)} & \multicolumn{1}{c}{$Q_{\rm{AME}}$ (keV)} & \multicolumn{1}{c}{(keV)}\\
\hline
I & $\ 861.955(47)$ & $\ 861.893(71)$ & 0.062(85)\\
II & $\ 861.965(27)$ & $\ 861.893(71)$ & 0.072(76)\\
Avg. & $\ 861.963(23)$ & $\ 861.893(71)$ & 0.070(75)\\
\end{tabular}
\end{ruledtabular}
\label{Table:Q_values}
\end{table}

\textit{Conclusion}---We have performed the first direct measurement of the $^{7}$Be electron capture $Q$-value using Penning trap mass spectrometry. The measured $Q$-value, $Q_{EC}$ = 861.963(23) keV, improves the precision in this quantity by a factor of three and is in agreement at the 1$\sigma$ level with the calculated value obtained using the masses of $^{7}$Be and $^{7}$Li listed in the most recent atomic mass evaluation.  The 23 eV uncertainty in the $Q$-value corresponds to a ~3.0 meV uncertainty in the $^{7}$Li recoil energy following $^7$Be EC decay, which was determined to be $T_{D}$ = 56.836(3) eV. A precise and accurate determination of the recoil energy is important for the BeEST experiment that has performed a precise measurement of the $^{7}$Li recoil spectrum to search for signatures of neutrino-coupled BSM physics. Our result will contribute to the evaluation of systematics in the BeEST experiment or to the validation of a positive result if such a signature is observed.  

Future work with $^{7}$Be EC in STJs could lead to sub-meV statistical and systematic uncertainties, requiring an improved measurement of $Q_{EC}$ to a precision of 1 eV or below. An order of magnitude increase in precision compared to the current measurement could be readily achieved using the phase imaging ion cyclotron resonance (PI-ICR) technique~\cite{Eliseev2013_PI-ICR,Eliseev2014_PI-ICR}, and further improvements could be made with a Penning trap that uses the image charge detection method e.g.~\cite{Myers2015_HeT,Rainville2004_MIT,Rau2020_LIONTRAP,Filianin2021_187Re}.

\section*{Acknowledgments}

This material is based upon work supported by the US Department of Energy, Office of Science, Office of Nuclear Physics under Awards No. DE-SC0015927, DE-SC0022538, DE-SC0021245 and DE-FG02-93ER40789. Support was provided by the National Science Foundation under Contracts No. PHY-1565546 and No. PHY-2111185, by Michigan State University and the Facility for Rare Isotope Beams, and by Central Michigan University.  KGL is also supported by the Gordon and Betty Moore Foundation (10.37807/GBMF11571).

\bibliography{MR_Refs.bib, ref.bib} 

\end{document}